\documentclass{amsproc}

\newtheorem{proposition}{Proposition}
\numberwithin{equation}{section}
\def\one{\mathbf{1}}

\begin{document}

\title{Integrable models and star structures}

\author{Ariel Garc\'{\i}a}
\address{Centre de Physique Th\'eorique, Luminy, 13288 Marseille, France}
\curraddr{Centro At\'omico Bariloche e Instituto Balseiro,
          8400 Bariloche, Argentina.}
\email{ariel@cab.cnea.gov.ar}

\author{Roberto Trinchero}
\address{Centro At\'omico Bariloche e Instituto Balseiro,
         8400 Bariloche, Argentina.}
\email{trincher@cab.cnea.gov.ar}
\thanks{A. G. was supported by CNRS, and R. T. by CONICET}

%%%%%%%%%%%%%%%%%%%%%%%%%%%%%%%%%%%%%%%%%%%%%%%%%%%%%%%%%%%%%%%%%%%%%%%%%%

\begin{abstract}
We consider the representations of Hopf algebras involved in some
physical models, namely, factorizable $S$-matrix models (FSM's),
one-{di\-men\-sio\-nal} quantum spin chains (QSC's) and statistical
vertex models (SVM's). These physical representations have definite
hermiticity assignments and lead to star structures on the corresponding
Hopf algebras. It turns out that for FSM's and the quantum mechanical
time-evolution of QSC's the corresponding stars are compatible with the
Hopf structures. However, in the case of statistical models the resulting
star structure is not a Hopf one but what we call a twisted star. Real
representations of a twisted star Hopf algebra do not close under the
usual tensor product of representations. We briefly comment on the
relation of these results with the Wick rotation.
\end{abstract}

\maketitle

%%%%%%%%%%%%%%%%%%%%%%%%%%%%%%%%%%%%%%%%%%%%%%%%%%%%%%%%%%%%%%%%%%%%%%%%%%

\section{Preliminary remarks}

This paper is devoted to the study of the $*$-structures on Hopf algebras
provided by certain families of physical models. The methods and presentation
used in describing these systems are the ones usually employed in theoretical
physics. However, we believe that the nature of the investigation and the
conclusions obtained are of interest to mathematicians, and could trigger
some interesting work from the mathematical side.

%%%%%%%%%%%%%%%%%%%%%%%%%%%%%%%%%%%%%%%%%%%%%%%%%%%%%%%%%%%%%%%%%%%%%%%%%%

\section{Factorizable $S$-matrix models and quantum groups}

\subsection{FSM's}
\label{FSM}

FSM's are one of the physical models where the quantum Yang-Baxter equation
first appeared in the physics literature \cite{Y}. By now there are a number
of books and reviews that deal with these systems \cite{zam,3}... In this
subsection we will briefly review some basic facts about them, however the
reader is referred to the above quoted literature for a complete treatment
of this subject.

The idea is to describe some particular scattering processes of particles
moving in one spatial dimension ($\mathbb{R}$). Such processes assume the
existence of asymptotically free particles in the initial (time
$t \to -\infty$) and final ($t \to +\infty$) states of the system.
Such quantum mechanical free particle states are described
by elements on a one-particle Hilbert space ${\bf H}^{(1)}$.
We denote the state of a free particle with rapidity\footnote{
If we deal with relativistic particles of mass $m$ in $1+1$
space-time dimensions, the momentum $p$ and the energy $E$ are
given in terms of the rapidity by
\begin{eqnarray}
p &=& m \sinh (\theta) \nonumber \\
E &=& m \cosh (\theta) \label{dr}
\end{eqnarray}
Note that Eqs. (\ref{dr}) are simply a parametrization of the
relativistic energy-momentum relation $E=\sqrt{p^2 + m^2}$.
}
$\theta \in \mathbb{R}$ and internal indices $i=1,\cdots, I$ by
$|i,\theta \rangle$.
Furthermore, one assumes that the states ${|i,\theta\rangle}$ form an
orthonormal and complete basis of ${\bf H}^{(1)}$. The inner product in
this basis given by
\begin{equation}
\label{1}
\langle i,\theta | j,\theta' \rangle
  = \delta_{ij} \, \delta(\theta - \theta') \ .
\end{equation}
The Hilbert space ${\bf H}^{(n)}$ for $n$ free particles is obtained as the
$n^{th}$ tensor product of the space ${\bf H}^{(1)}$, the inner product in
${\bf H}^{(n)}$ being the extension of (\ref{1}) to the tensor product.

It will be useful for our purely algebraic purposes to think ${\bf H}^{(1)}$
as giving an $I$-dimensional complex vector space $V(\theta)$ for each value
of the rapidity $\theta$.

Let $B_{ij}^{kl}(\theta )$ be the probability amplitude\footnote{
This means that the probability of such a process is given by the modulus
squared of the probability amplitude $B_{ij}^{kl}(\theta )$.
}
for the scattering of 2 particles, of types $i$ and $j$, into two particles
of types $k$ and $l$. If the initial states have rapidities
$\theta_1,\theta_2$, then the scattered particles will have the same (but
interchanged) rapidities, and the scattering amplitude $B_{ij}^{\,kl}$
will only depend on the difference $\theta_1 - \theta_2 \equiv \theta$,
due to Poincar\'e invariance. Such probability amplitude corresponds to the
following matrix element of the unitary evolution operator ${\hat S}$ of the
model:
\begin{equation}
\label{2}
   B_{ij}^{\,kl}(\theta ) \, \delta(\theta + {\theta}')
   = \langle i,j,\theta |\hat{S}|k,l,{\theta}' \rangle \ .
\end{equation}
Factorizability of the model means that any many-particle scattering
amplitude can be written as the product of two-to-two particles scattering
amplitudes. Unicity of this factorization requires the following identity
for the scattering of three particles:
\begin{equation}
\label{YB}
B_{12}(\theta _{23})B_{23}(\theta _{13})B_{12}(\theta _{12})
=B_{23}(\theta _{12})B_{12}(\theta _{13})B_{23}(\theta _{23})
\end{equation}
where $\theta _{ab}=\theta _{a}-\theta _{b}$ are the rapidity differences
by pairs of the three particles involved in the equality (\ref{YB}). Hence
they are not independent, and satisfy
\begin{equation}
\theta_{12} = \theta_{13} - \theta_{23} \ .
\end{equation}
The subindices of $B$ denote its action on the vector space
$V(\theta _{1}) \otimes V(\theta _{2}) \otimes V(\theta _{3})$.
We define the $R$-matrix by
\begin{equation}
R(\mu )=B(\mu ) P \ ,
\end{equation}
where $P$ is the permutation operator. Equation (\ref{YB}) can be
rewritten in terms of $R$ as
\begin{equation}
\label{ybr}
R_{12}(\lambda -\mu )R_{13}(\lambda )R_{23}(\mu ) =
R_{23}(\mu )R_{13}(\lambda )R_{12}(\lambda -\mu ) \ ,
\end{equation}
where $\lambda =\theta _{13}\;, \; \mu =\theta _{12}$. This last equation,
or also Eq. (\ref{YB}), is called the Quantum Yang-Baxter equation with
spectral parameter. All of the above comes directly from the theory of FSM's
and nothing about QG's is employed.

\subsection{QG's with spectral parameter}
\label{sect:2.2}

Consider a set of elements $T_i^j(\lambda)$, for every value of the
parameter $\lambda \in \mathbb{C}$ and $i,j=1,\cdots,I$. The free algebra
they generate can be turned into a bialgebra $\mathcal{F}_o$ by defining the
coproduct as $\Delta T_i^j(\lambda) = T_i^k(\lambda)\otimes T_k^j(\lambda)$
and the counit as $\epsilon (T_i^j(\lambda)) = \delta_i^j$.
Let us now consider a collection of vector spaces $V(\lambda)$ for
every $\lambda \in \mathbb{C}$, with o.n. basis
$\{e_i(\lambda ) \,,\;i=1,\cdots,I\}$, and take the linear map
\begin{equation}
\label{blambda}
\begin{array}{rcl}
B(\lambda -\mu )\: : \; V(\lambda )\otimes V(\mu )
 & \longrightarrow & V(\mu )\otimes V(\lambda ) \\
 && \\
e_{i}(\lambda )\otimes e_{j}(\mu ) &\longrightarrow &
B_{ij}^{kl}(\lambda -\mu )\, e_{k}(\mu )\otimes e_{l}(\lambda ) \ ,
\end{array}
\end{equation}
to be a spectral parameter dependent solution of the Yang-Baxter equation
(\ref{YB}). It is not difficult to show that taking the quotient of
${\mathcal{F}}_o$ by the two-sided ideal $I(R)$ generated by the
expression
\begin{equation}
R: \quad R_{ij}^{kl}(\lambda -\mu)\; T_k^p(\lambda )\,T_l^q(\mu ) =
  T_{j}^{l}(\mu )\, T_{i}^{k}(\lambda )\, \, R_{kl}^{pq}(\lambda -\mu )
\label{1.12}
\end{equation}
you get a bialgebra\footnote{
\textbf{Comodule algebra and Zamolodchikov's operators} Let us
consider the quantum linear space defined by the quadratic algebra given
by the quotient
$$
A=\frac{\langle V(\lambda)\rangle_\lambda}{R}
$$
of the free algebra $\langle V(\lambda)\rangle_\lambda$ by the
relation
$$
R \:: \;\; \cup_{\lambda ,\mu} \;
[1\otimes 1-B(\lambda -\mu)]\, \, \, V(\lambda)\otimes V(\mu) \ .
$$

An ${\mathcal{F}}$-comodule structure on $V(\lambda)$ is given by
the coaction
$\delta _{V(\lambda )}\, e_{i}(\lambda ) =
T_{i}^{k}(\lambda )\otimes e_{k}(\lambda )$.
The definition of the coaction on a tensor product
$V(\lambda )\otimes V(\mu )$ is
$$
\delta _{V(\lambda )\otimes V(\mu )}=
(m\otimes I_{V(\lambda)}\otimes I_{V(\mu )})
(I_{\mathcal{F}}\otimes \tau \otimes I_{V{(\mu )}})
(\delta _{V(\lambda )}\otimes \delta _{V(\mu )}) \ ,
$$
where $\tau $ is the flip operator. $A$ is the algebra obeyed by
Zamolodchikov's operators \cite{zam}, or the also called spectral
parameter dependent quantum plane algebra. Note that this comodule action
preserves the quadratic relations of the algebra $A$, i.e.,
$$
\left( I_{\mathcal{F}}\otimes R\right) \delta_{V\otimes V} =
\delta_{V\otimes V}\, R \ .
$$
}
${\mathcal{F}}\equiv {\mathcal{F}}_o/{I(R)}$ \cite{1,cp}. Note that $I(R)$
is both an ideal and a coideal, and that
$R_{ij}^{kl}(\mu)=B_{ij}^{lk}(\mu)$.

\subsection{Relation between FSM's and QG's with spectral parameter}

Using the 2-2 $S$ matrix of (\ref{2}) we can build the following
representation of the bialgebra of the previous subsection:
\begin{eqnarray}
\rho : \: {\mathcal{F}} & \rightarrow  & T(V(\lambda )) \nonumber \\
\rho (T_i^k(\lambda ))_j^l & = & R_{ij}^{kl}(\lambda ) \ .
\label{rep}
\end{eqnarray}
Replacing (\ref{rep}) into (\ref{1.12}) we reobtain (\ref{ybr}), showing
that it is indeed a representation. Note that from the point of view of
FSM's, this representation (\ref{rep}) is such that
\begin{equation}
R_{ij}^{kl}(\theta ) \delta(\theta+{\theta}') =
   B_{ij}^{lk} (\theta) \delta(\theta+{\theta}') =
   \langle i,j,\theta |{\hat{S}}|l,k,{\theta}' \rangle \ .
\end{equation}
Moreover, all the $T_i^j(\lambda)$ commute with the scattering matrix
$\hat{S}$, as operators on the $n$ particle Hilbert space given by the
(tensor product) representation $\rho$:
$$
   \rho_{_\otimes}[T_i^j(\lambda)] \, \hat{S} =
   \hat{S} \, \rho_{_\otimes}[T_i^j(\lambda)] \ , \quad \forall \lambda,i,j
$$
This corresponds, physically, to an additional scattering with a particle of
rapidity $\lambda$ and (in, out) indices $i,j$.

\subsection{Stars}
\label{subsect:FSMstars}

Since we have on $\textbf{H}^{(n)}$ the (positive definite) scalar product
given in subsection \ref{FSM}, we have the notion of the adjoint of an
operator. The physical requirement of unitarity of the ${\hat{S}}$ matrix
operator\footnote{
This requirement is at the roots of quantum mechanics, and
it is a direct consequence of the structure of Schr\"odinger's equation.
}
together with above mentioned definition of the adjoint leads to a unitary
$B$ matrix. Furthermore, if one assumes $\rho$ to be a star representation
of the bialgebra ${\mathcal{F}}$ with a star structure, then one can
determine this star on ${\mathcal{F}}$. From the discussion above, we have
\begin{equation}
\label{241}
  B_{ij}^{lk}(\theta)(B^{\dagger})_{lk}^{mn}(\theta) =
    \delta _{i}^{m}\delta _{j}^{n} =
  (B^{\dagger})_{ij}^{lk}(\theta)B_{lk}^{mn}(\theta) \ .
\end{equation}
Using Eq. (\ref{rep}), and considering that
$$
  (B^\dagger)_{ij}^{lk} = \overline{R_{lk}^{ji}}
    = [\rho(T_l^j)^\dagger]_i^k \ ,
$$
we may rewrite (\ref{241}) as
\begin{eqnarray}
B_{ij}^{lk}(\theta ) \: (B^\dagger)_{lk}^{mn}(\theta)
   &=& \rho[T_i^k(\theta)]_j^l \: \rho[(T_m^k(\theta))^*]_l^n
    = \delta _i^m \, \delta_j^n \ , \nonumber \\
(B^\dagger)_{ij}^{lk}(\theta ) \: B_{lk}^{mn}(\theta )
   &=& \rho[(T_l^j(\theta))^*]_i^k \: \rho[T_l^n(\theta )]_k^m
    = \delta _i^m \,\delta_j^n \ .
\end{eqnarray}
These last equalities can be fulfilled if
\begin{eqnarray}
   T_i^k(\theta)\, \left[ T_j^k(\theta)\right]^* &=&
        \delta_{ij}\:\one \nonumber \\
   \left[T_k^i(\theta)\right]^*\, T_k^j (\theta) &=& \delta^{ij}\:\one \ .
\label{242}
\end{eqnarray}
Therefore the algebraic structure provided by this formulation of the
FSM's is the one involved in the following proposition.

\begin{proposition}
\label{p1}
Let $A$ denote the $*$-bialgebra generated by ${\mathbf{1}}$ and the
$T_i^j(\theta)$ ($i,j=1,\cdots,N ; \:\theta \in \mathbb{R}$) satisfying
the relations
\begin{eqnarray}
   R_{ij}^{kl}(\lambda - \mu) \; T_k^p(\lambda) \, T_l^q(\mu)
   &=& T_j^l(\mu)\, T_i^k(\lambda ) \; R_{kl}^{pq}(\lambda - \mu)
        \label{243} \\
   T_i^k(\theta)\, \left[T_j^k(\theta)\right]^* &=& \delta_{ij} \: \one
   \label{243'} \\
   \left[T_k^i(\theta)\right]^* \, T_k^j(\theta) &=& \delta^{ij} \: \one
   \label{243''} \ ,
\end{eqnarray}
where $R$ is a solution of the Yang-Baxter equation (\ref{ybr}). The
coproduct and counit are taken to be $*$-homomorphisms,
\begin{eqnarray*}
\Delta(a^*) &=& [\Delta a]^* \\
\epsilon(a^*) &=& \overline{\epsilon(a)} \;\;,\quad a \in A \ ,
\end{eqnarray*}
given on the generators by
\begin{align}
\label{244}
   \Delta(T_i^j(\theta)) &= T_i^k(\theta) \otimes T_k^j(\theta) \ , &
       \Delta(\one) &= \one \otimes \one \\
   \epsilon (T_i^j(\theta)) &= \delta_i^j \ , &
       \epsilon(\one) &= 1 \ . \nonumber
\end{align}
For the star structure on the tensor product we consider two possibilities,
\begin{enumerate}
\item[i)] $(a \otimes b)^* = a^* \otimes b^* \;\;;\quad a,b \in A$
           (Hopf star),
\item[ii)] $(a \otimes b)^* = b^* \otimes a^* \;\;;\quad a,b \in A$
           (twisted star).
\end{enumerate}
Furthermore, we consider a comodule $V(\theta)$ ($\dim(V(\theta)) = I$)
for each value of $\theta$. Let $\{ e_i(\theta)\;;\quad i=1,\cdots,I\}$
be a basis of $V(\theta)$, and define a right coaction on $V(\theta)$ by
\begin{equation}
\label{245}
\delta e^i(\theta) = e^j(\theta) \otimes T_j^i(\theta) \ .
\end{equation}
Then:
\begin{enumerate}
\item The linear antihomomorphism $S$ defined on the generators by
\begin{equation}
\label{246}
   S(T_i^j(\theta)) = \left[T_j^i(\theta) \right]^* \;\;, \quad
   S(\one) = \one \ ,
\end{equation}
is an antipode for the algebra $A$, which therefore has a Hopf algebra
structure.
\item As $A$ is non-cocommutative ($N \ge 2$), only possibility $(i)$
for the star closes without requiring additional relations.
\item The scalar product on $V= \oplus_{\theta} V(\theta)$ determined by
\begin{equation}
\label{247}
(e^i(\theta), e^j(\theta')) = \delta^{ij} \delta(\theta - \theta')
\end{equation}
is such that the coaction (\ref{245}) induces a $*$-representation
of the dual\footnote{
The (Hopf) star structure on $A$ induces a (Hopf) star on the
dual $\tilde{A}$, which is obtained from
\begin{equation}
\label{247'}
   \langle h^* , a \rangle = \overline{\langle h, [Sa]^* \rangle} \ ,
\end{equation}
where $a \in A$, $h \in \tilde{A}$ and $\langle\:,\:\rangle$ is the
duality pairing between $A$ and $\tilde{A}$.
}
$\tilde{A}$ of $A$ on $V$. This scalar product is invariant under
the action of $\tilde{A}$ in the sense of reference \cite{cgt}.
\end{enumerate}
\end{proposition}

We refer the reader to \cite{cgt} for a detailed discussion of
twisted and Hopf stars and some of their properties. The
proof of this Proposition is an easy calculation. For point $(1)$
it suffices to show that $S$ satisfies the axioms of the antipode on
the generators, which involves essentially relations (\ref{243'}) and
(\ref{243''}). Regarding statement $(2)$, applying $\Delta$ to (\ref{243'})
and assuming $*$ to be a Hopf star, one sees that the obtained relation is
automatically satisfied. On the other hand, in the twisted star case new
(additional) relations between the $T_i^j\:$'s need to be true in order to
have a consistent $*$-bialgebra. The proof of $(3)$ is a simple calculation
that again involves in an essential way relations (\ref{243'})
and (\ref{243''}).

In fact, the proof of point $(1)$ can be reversed and used to show that
if one starts with an $RTT$ {\em Hopf} algebra (an $RTT$ bialgebra as
above plus an antipode $S$), a Hopf star $*$ satisfying (\ref{243'}) and
(\ref{243''}) can be immediately defined as an antilinear antimorphism by
\begin{equation}
\label{248}
   \left[T_i^j(\theta)\right]^* = S(T_j^i(\theta)) \;\;, \quad
   \one^* = \one \ .
\end{equation}
Hence, we conclude that at least for a certain class of FSM's (for
instance, all those described by an $R$-matrix of the $GL_q(N)$ type)
the real structures on the underlying Hopf algebras are Hopf stars. In
other words, the combined requirement of having a quantum group symmetry
and the physical unitarity of the system (plus some naturalness
assumptions, as the star representation condition) restrict the $RTT$
bialgebra to be a true Hopf algebra. An example of a field theoretic model
that leads to a factorizable $S$-matrix is given by the sine-Gordon model.
The scattering of solitonic and antisolitonic asymptotic states in this
model is described by a factorizable and unitary $S$-matrix \cite{zam,3},
associated to a quantum group of $SL_q$ type.

%%%%%%%%%%%%%%%%%%%%%%%%%%%%%%%%%%%%%%%%%%%%%%%%%%%%%%%%%%%%%%%%%%%%%%%%%%

\section{Integrable quantum spin chains. The $XXZ$-model}

In this section we consider the case of integrable one dimensional spin
chains. All the reasoning will be exemplified by considering the
$XXZ$-model, however most of the steps are quite general and apply to any
one dimensional spin chain. We are interested in the representation of the
underlying quantum group that this kind of models provide. We choose to make
this representation explicit by obtaining a two dimensional classical vertex
model out of the quantum spin chain. Such relation between $d$-dimensional
quantum systems and $(d+1)$-dimensional classical systems is quite general
\cite{suz} and is essentially what is often referred in physics as the path
integral formulation of quantum mechanics \cite{fey}.

Let us consider a linear lattice of $N$ sites labelled by an index
$k=1, \cdots ,N$. To each site $k$ of this lattice we associate a complex
$n$-dimensional vector space ${\bf H}_{(k)}$ ($n=2$ for the spin $1/2$
$XXZ$ case). In each of these spaces ${\bf H}_{(k)}$ we consider an
irreducible $n$-dimensional representation $\sigma_a$ of the generators
of the Lie algebra of $SU(2)$. We choose them so as to satisfy the
following algebraic relations:
\begin{eqnarray}
\label{2.1}
[\sigma_a , \sigma_b ] &=& 2 i \epsilon_{abc} \sigma_c \ , \\
(\{ \sigma_a , \sigma_b \} &=& 2 \delta_{ab} \ , \quad
   \hbox{for the spin $1/2$ case}) \nonumber
\end{eqnarray}
where $a,b,c=1,2,3$, $\epsilon_{abc}$ is the totally antisymmetric
tensor with $\epsilon_{123} = 1$, and $\{\cdot , \cdot \}$ stands for
the anticommutator. The total vector space of the chain is taken to be
$\mathbf{H} = \bigotimes_k \mathbf{H}_{(k)}$. We define spin operators
acting on this space by 
{\small
$\sigma_a (k) = 1\otimes \cdots \otimes 1 \otimes 
   \stackrel{k}{\overbrace{\sigma_a}} \otimes 1\otimes\cdots\otimes 1$
}.
The Hamiltonian is also an operator acting on $\mathbf{H}$. For the 
case of the spin $1/2$ $XXZ$-model, it is given by
\begin{align}
\label{2.4}
   H &= \sum_{k=1}^{N} H_{k,k+1} \ , \\
\label{2.5}
H_{k,k+1} &= \sigma_1(k)\,\sigma_1(k+1) + \sigma_2(k)\,\sigma_2(k+1) +
            J \,\sigma_3(k)\,\sigma_3(k+1) \ .
\end{align}
In the above formula for $H$ we impose periodic boundary conditions, i.e.,
\begin{equation}
\label{2.6}
\sigma_a(N+1) = \sigma_a(1) \ , \quad \forall a \ .
\end{equation}
Note that $[H_{k,k+1}, H_{j,j+1}] \neq 0 \;$ only if $\;j=k+1\;$ or
$\;k-1$.

The quantity of physical interest is the operator
\begin{equation}
\label{2.7}
   U = \exp{[-zH]} \ .
\end{equation}
If we are doing quantum mechanics, we take $z=it$ ($t \in \mathbb{R}$ is the 
time) and $U$ is called the time-evolution operator. If we are doing
statistical mechanics of the QSC, instead, we take $z=\beta=1/(k_B T)$
the inverse temperature ($z \in \mathbb{R}$), and $U$ would be the quantum
Boltzmann operator of the chain. Using the parameter $z$ we can analyse
simultaneously both cases. Now we may rewrite $U$ using Trotter's
formula, which is valid for bounded operators \cite{trot},

\begin{equation}
\label{2.8}
\exp{\sum_i A_i} = \lim_{L \to \infty}
   \left[ \prod_i \exp{ \frac{1}{L} A_i} \right]^L \ .
\end{equation}
Hence we have
\begin{align}
\label{2.11}
   U &= \lim_{L \to \infty} T(\epsilon)^L \ , \\
\intertext{where $T(\epsilon)$ is the transfer matrix}
\label{2.9}
T(\epsilon) &= \prod_{k=1}^{N} B_{k,k+1} \\
\label{2.10}
B_{k,k+1} &= \exp{\left[ -\epsilon H_{k,k+1}\right]} \ , \quad
   \hbox{with} \quad \epsilon = \frac{z}{L} \ .
\end{align}
Therefore we are led to the evaluation of matrix elements of the
transfer matrix. In order to do so we choose an orthonormal basis of
the ${\bf H}_{(k)}$ that we denote $\{ |j\rangle \}\:, \; j=1,\cdots ,n$.
Having a basis for each ${\bf H}_{(k)}$, we construct a basis
$\{ |i_1 , \cdots , i_N \rangle \}$ of ${\bf H}$ by taking the tensor
product, $|i_1 , \cdots , i_N \rangle =
    |i_1 \rangle \otimes \cdots \otimes |i_N \rangle$.
Note that the operators $B_{k,k+1}$ only act non trivially on the states
of sites $k$ and $k+1$. Thus the only factor of $T(\epsilon)$ that
acts non-trivially on the vector space corresponding to site $k$ is the
product of operators $B_{k-1,k}\,B_{k,k+1}$.

We consider next the matrix element
\begin{equation}
\label{2.13}
\langle i_1 , \cdots , i_N| T(\epsilon) |j_1 , \cdots , j_N \rangle =
\langle i_1 , \cdots , i_N|
        \prod_{k=1}^N B_{k,k+1} |j_1 , \cdots , j_N \rangle \;\;.
\end{equation}
One could now introduce an identity of the form
$\one = \sum_{p_k} |p_k \rangle \langle p_k |$ between the operators
$B_{k-1,k}\,B_{k,k+1}$. However, doing so would break explicit
translational invariance along the chain, because the sites $1$ and $N$
are treated differently. This problem can be solved if one remarks that
Trotter's formula (\ref{2.8}) would still be valid if we keep only first
order terms in $1/L$ inside the square brackets. This means that only
$\mathcal{O}(\epsilon)$ terms matter in (\ref{2.9}). Considering this,
it is easy to see that the matrix element (\ref{2.13}) of the infinitesimal
``evolution'' (in time or temperature) operator may be expanded in a
translation-invariant way as
\begin{eqnarray}
\label{2.14}
\langle i_1 , \cdots , i_N| T(\epsilon) |j_1 , \cdots , j_N \rangle
   &=& B_{p_1,i_2}^{j_1,p_2}(\epsilon) \, B_{p_2,i_3}^{j_2,p_3}(\epsilon) \,
       B_{p_3,i_4}^{j_3,p_4}(\epsilon) \cdots \\
   & & \quad B_{p_{N-1},i_N}^{j_{N-1},p_N}(\epsilon) \,
       B_{p_N,i_1}^{j_N,p_1}(\epsilon)
     + \mathcal{O}(\epsilon^2) \ . \nonumber
\end{eqnarray}
Here we made use of the notation
\begin{equation}
\label{2.15}
  B_{ij}^{kl}(\epsilon) \equiv \langle i,j| B(\epsilon) |k,l\rangle \ ,
\end{equation}
for each (any) pair of particles. In fact, the $\mathcal{O}(\epsilon^2)$
terms in (\ref{2.14}) become irrelevant in the $N \to \infty$ limit, as
they are associated to the chosen ``endpoints'' of the chain.

Being $P$ the permutation map on
$\mathbf{H}_{(k)} \otimes \mathbf{H}_{(k+1)}$, we introduce the operator
$R(\epsilon) = B(\epsilon) \, P$, so
\begin{equation}
\label{2.15'}
  R_{ij}^{kl} = B_{ij}^{lk} \ .
\end{equation}
Now (\ref{2.14}) reads
\begin{eqnarray}
\label{2.16}
\langle i_1 , \cdots , i_N| T(\epsilon) |j_1 , \cdots , j_N \rangle
   &=& R_{p_1,i_2}^{p_2,j_1}(\epsilon) \, R_{p_2,i_3}^{p_3,j_2}(\epsilon) \,
       R_{p_3,i_4}^{p_4,j_3}(\epsilon) \cdots \\
   & & \quad R_{p_{N-1},i_N}^{p_N,j_{N-1}}(\epsilon) \,
       R_{p_N,i_1}^{p_1,j_N}(\epsilon)
     + \mathcal{O}(\epsilon^2) \ . \nonumber
\end{eqnarray}

If we now assume that this $R$-matrix satisfies Yang-Baxter equation (this
will be proved later for the spin $1/2$ $XXZ$ model), we may use it to
define an $RTT$ bialgebra through equation (\ref{1.12}) as in
section~\ref{sect:2.2}. As before, the $R$-matrix itself is a
representation of this bialgebra,
\begin{equation}
\label{2.17'}
   \langle i |T_a^b(\lambda) |j\rangle =
      \rho (T_a^b(\lambda))_i^j = R_{ai}^{bj}(\lambda) \ .
\end{equation}

Therefore, the transfer matrix of the spin model can be written in terms
of the trace $T_a^a(\lambda) \equiv \sum_a T_a^a (\lambda)$ (trace over
the auxiliary space) of the $T_a^b$ operators:
\begin{eqnarray}
\label{2.17}
\langle i_1, \cdots , i_N| T |j_1, \cdots , j_N \rangle
   &=& \langle i_2, \cdots , i_N,i_1|
           \rho(T_{p_1}^{p_2})\, \rho(T_{p_2}^{p_3}) \cdots \\
   & & \quad\qquad \rho(T_{p_{N-1}}^{p_N})\, \rho(T_{p_N}^{p_1})
       |j_1, \cdots , j_N \rangle \nonumber \\
   &=& \langle i_1, \cdots , i_N| C^\dagger \rho_{_\otimes}(T_p^p)
       |j_1, \cdots , j_N \rangle \ . \nonumber
\end{eqnarray}
Here $C$ is the (unitary) cyclic permutation operator,
$$
C \, |j_1, j_2, \cdots , j_N \rangle = |j_2, \cdots , j_N, j_1 \rangle \ .
$$
An easy calculation tells us that\footnote{We have dropped the $\rho$'s
from the formulas.
}
$$
\left[ C, T_a^a(\mu) \right] = 0 \ , \quad \forall \mu \ ,
$$
which is not true for arbitrary $T_a^b$'s. As the $T_a^a(\mu)$ commute for
different values of the parameter, this implies that we have an infinite
set (parametrized by $\mu$) of conserved quantities:
$$
   \left[ U(z), T_a^a(\mu) \right] = 0 \ , \quad \forall \mu \ .
$$

Coming back to the $R$ matrix, we will now show that in the spin $1/2$ case
of the $XXZ$-model, it can be chosen as a solution of the YB equation.
We are interested only in the first order terms in $R$. Using (\ref{2.10})
we easily get
{\small
\begin{equation}
\label{2.20}
||R||_{ij}^{kl}(\epsilon) =
   \left(\begin{array}{cccc}
         1-J\epsilon & 0 & 0 & 0 \\
         0 & -2\epsilon & 1+J\epsilon & 0 \\
         0 & 1+J\epsilon & -2\epsilon & 0 \\
         0 & 0 & 0 & 1-J\epsilon
   \end{array}\right) + \mathcal{O}(\epsilon^2) \ .
\end{equation}
}

This matrix is of the ``six-vertex model'' type \cite{3}. Solutions of
this form to the Yang-Baxter equation (\ref{ybr}) exist and can be
parametrized as
{\small
\begin{equation}
\label{2.20'}
||R_\alpha||_{ij}^{kl}(u) =
   \left(\begin{array}{cccc}
         \sinh(u+\alpha) & 0 & 0 & 0 \\
         0 & \sinh(u) & \sinh(\alpha) & 0 \\
         0 & \sinh(\alpha) & \sinh(u) & 0 \\
         0 & 0 & 0 & \sinh(u+\alpha)
   \end{array}\right) \ .
\end{equation}
}

Remark now that the YB equation is not reparametrization invariant, and
$\epsilon$ could not be the ``right'' parameter to get a given $R$ matrix
to satisfy (\ref{ybr}). Note also that if $R(\lambda)$ is a
solution of the Yang-Baxter equation then $f(\lambda) R(\lambda)$ is also
a solution for any scalar function $f$ of the parameter $\lambda$.

Taking into account the above remarks, we see that the matrices $R$ and
$R_\alpha$ of equations (\ref{2.20}) and (\ref{2.20'}) coincide, up to
a global factor and $\mathcal{O}(\epsilon^2)$ terms, if we take
\begin{eqnarray}
\label{epsilon-u}
   \epsilon(u) &=& -\frac{1}{2} \frac{\sinh u}{\sinh \alpha} \\
   J &=& \cosh \alpha \ .
\end{eqnarray}
If we are considering the quantum mechanical time-evolution of
the QSC, remembering that $\epsilon = it/L \in i \,\mathbb{R}$ and 
$J \in \mathbb{R}$ we see that $u \in i \,\mathbb{R}$. In the statistical
case we have $u \in \mathbb{R}$ instead.

\subsection{Stars}

For the quantum mechanical time-evolution of any quantum integrable spin
chain the $B$-matrix we obtain is unitary, since it is given by matrix
elements in a Hilbert space of the unitary operator $\exp{(-itH/L)}$, cf.
equation (\ref{2.10}). Therefore, we could repeat exactly the same
argument of section~\ref{subsect:FSMstars} and conclude that it is natural
to endow the $RTT$ bialgebra introduced in (\ref{2.17'}) with a
$\Delta$-compatible (untwisted) star operation $*$. This is so in the
cases where the $R$-matrix allows the existence of an antipode for the
bialgebra, as shown by Proposition~\ref{p1}.

If we do a statistial study of the same QSC, now $z=\beta \in \mathbb{R}$, 
and the operator $B$ is hermitian. The analysis of star operations for
this case will be done in the next section, where we will consider SVM's.

%%%%%%%%%%%%%%%%%%%%%%%%%%%%%%%%%%%%%%%%%%%%%%%%%%%%%%%%%%%%%%%%%%%%%%%%%%

\section{Statistical vertex models}

We now consider a (classical) statistical vertex model defined for an
$N\times M$ square lattice ($N$ ``horizontal'' sites and $M$ ``vertical''
ones, and periodic boundary conditions). Each vertex in the lattice has a
Boltzmann weight $R_{ij}^{kl}(\beta) = \exp{\left[-\beta E(ij;kl)\right]}$
associated to it, corresponding to the energy of the configuration
($i,j,k,l$) of the four links converging to the vertex and to an inverse
temperature $\beta$. This matrix $R$ of termal weights is evidently real,
and would also be symmetric ($R_{ij}^{kl}=R_{kl}^{ij}$) if we assume the
energy of each configuration to be invariant under a diagonal reflection
of the vertex. This happens in the so called {\it six} and {\it eight
vertex models}. As we are talking about integrable systems, we also require
$R$ to satisfy YB equation (\ref{ybr}).

We refer the reader to a standard reference such as \cite{3} for a
detailed analysis of these vertex models. However, here we need at
least to sketch the way of relating them to Quantum Groups to be able
to introduce the physically relevant stars.

The basic quantity in a statistical system is always the partition
function,
$$
  Z = \sum_{\mathrm{configurations}\ c}\exp{\left[ -\beta E(c)\right]} \ ,
$$
as it encodes all the physical information of the system. In this
particular case $Z$ may be rewriten as
\begin{eqnarray*}
   Z &=& \sum_{\mathrm{configurations}\ c\;\;}
         \prod_{\mathrm{vertices}\ v} R(c[v]) \ .
\end{eqnarray*}
Here we can group the Boltzmann weights by rows to build transfer matrices
(compare with Eq.(\ref{2.16})\ldots)
\begin{eqnarray}
\label{transf-mat}
   t_{i_1,\cdots,i_N}^{j_1,\cdots,j_N} &\equiv&
      R_{b_1 i_1}^{b_2 j_1} \, R_{b_2 i_2}^{b_3 j_2} \cdots
      R_{b_N i_N}^{b_1 j_N} \ ,
\end{eqnarray}
which may then be thought as matrix elements on an $N$-site Hilbert space
of a self adjoint operator (assuming that $R$ is a real, symmetric matrix).
Moreover, using $R$ we can introduce an $RTT$ bialgebra exactly as in the
previous sections, and so the transfer matrices (\ref{transf-mat}) happen to
be matrix elements of $T_a^a$ in an $N^{th}$ tensor product representation.
In fact, what we have done in the section about QSC's, was to rewrite
the evolution of the quantum system in terms of transfer matrices
(\ref{transf-mat}) of a (classical) statistical system\ldots Here all the
same formulas apply, except that now the partition function does not include
the cyclic permutation operators $C$ that we previously found in $U$, and
that $\epsilon = \beta \in \mathbb{R}$.

As a consequence, if we assume the representation given by
Eq. (\ref{2.17'}) to be a star representation of a $*$-algebra defined
by Eq. (\ref{1.12}), then we can determine this star by writing
\begin{eqnarray}
\label{2.27}
\langle i| \rho[T_a^b(\mu)] |j\rangle &=& R_{ai}^{bj}(\mu)
    = \overline{R^{ai}_{bj}(\mu)} \\
   &=& \langle i| \rho[T_b^a(\mu)]^{\dagger} |j\rangle
    = \langle i| \rho[(T_b^a(\mu))^*] |j\rangle \ , \quad \forall i,j \ .
      \nonumber
\end{eqnarray}
From this we obtain
\begin{equation}
\label{2.28}
   \left[ T_a^b(\mu) \right]^* = T_b^a(\mu) \ ,
\end{equation}
as a sufficiency condition. Therefore, the SVM's characterized by a
symmetric $R$ matrix fit in the hypothesis of the following proposition.

\begin{proposition}
\label{p2}
Under the hypothesis of Proposition~\ref{p1} but replacing (\ref{243'})
and (\ref{243''}) by
\begin{equation}
\label{2.29}
   \left[T_a^b(\theta) \right]^* = T_b^a(\theta) \ ,
\end{equation}
we have:
\begin{enumerate}
\item As the bialgebra $A$ is non-cocommutative for $N \ge 2$, only
possibility $(ii)$ (twisted star) for the $*$-structure on $A \otimes A$
is consistent.
\item The following scalar product on $V = \bigotimes_{\theta} V(\theta)$
fixed by
\begin{equation}
   (e^i(\theta), e^j(\theta')) = \delta^{ij} \delta(\theta - \theta')
\end{equation}
is such that the coaction (\ref{245}) induces a $*$-representation
of the dual $\tilde{A}$ of $A$ on $V$.
\end{enumerate}
\end{proposition}

\noindent
Hence we conclude that for SVM's (with a symmetric $R$), or for any model
with an hermitian $R$-matrix (as is the case of statistical QSC's with a
Hamiltonian symmetric in neighboring sites\footnote{
Even if the $B_{ij}^{kl}$ matrix is always hermitian, this is not true in
general for $R_{ij}^{kl}$ unless the property $R_{ij}^{kl}=R_{ji}^{lk}$
holds.
}),
the star operation on the underlying Hopf algebra is a twisted star.

It is interesting to note that if one performs a Wick rotation
($\epsilon \to i\epsilon$) on this statistical system then one obtains
an unitary $R$ matrix that corresponds, as we have seen for the FSM's,
to a Hopf star.

\subsection{Stars in $U_q(sl(2))$ and the star in SVM's}

The star defined by (\ref{2.28}) is not a Hopf star, but a twisted star
instead, for the bialgebra defined by Eq. (\ref{1.12}) and the coproduct
(\ref{244}). We have a classification of the Hopf stars in $U_q(sl(2))$
but we do not have a classification of the stars of the bialgebras
defined by relations (\ref{1.12}). However, both are related. For the
case of the $XXZ$-model this relation is given in \cite{4}. The
$R$-matrix of this model is given by (\ref{2.20'}), or, changing
variables ($u=i\gamma\delta$, $\alpha=i\gamma$) and basis, by
{\small
\begin{equation}
\label{17}
R(\lambda) = i\, \left(\begin{array}{cccc}
\sin \gamma (\delta +1) & 0 & 0 & 0\\
 0 & \sin \gamma \delta & \exp (-i\gamma \delta )\sin \gamma & 0\\
 & \exp (i\gamma \delta)\sin \gamma & \sin \gamma \delta & \\
 &  &  & \sin \gamma (\delta +1)
\end{array}\right) \ ,
\end{equation}
}
where $i\delta=\lambda$, $q=\exp i\gamma$.

Defining the operator valued matrices
\begin{equation}
\label{18}
L_+ = q^{\frac{1}{2}}\left( \begin{array}{cc}
k^{\frac{1}{2}} & (q-q^{-1})x_{-}\\
0 & k^{-\frac{1}{2}}
\end{array}\right) \ , \ L_- = q^{-\frac{1}{2}}\left( \begin{array}{cc}
k^{-\frac{1}{2}} & 0\\
-(q-q^{-1})x_{+} & k^{\frac{1}{2}}
\end{array}\right) \ ,
\end{equation}
we build using them the $T(\lambda)$ operators:
\begin{equation}
\label{19}
T(\lambda )=e^{\lambda \gamma }\, L_+ - e^{-\lambda \gamma }\, L_- \ .
\end{equation}
Replacing (\ref{19}) into (\ref{ybr}) you get an identity if
$x_+,x_-$ and $k$ satisfy the algebraic relations of $U_q(sl(2))$. The
$R$ matrix (\ref{17})gives a representation of this algebra: comparing
(\ref{18}) with (\ref{17}) it is given by
\begin{equation}
\label{20} x_{-}=\left( \begin{array}{cc}
0 & 0\\ q^{-\frac{1}{2}} & 0
\end{array}\right) ,\: x_{+}=\left( \begin{array}{cc} 0 & q^{\frac{1}{2}}\\
0 & 0 \end{array}\right) ,\: k=\left( \begin{array}{cc}
q & 0\\ 0 & q^{-1}
\end{array}\right) \ .
\end{equation}
Taking this to be a star representation of $U_q(sl(2))$ leads to the
following twisted star structure:
\begin{equation}
   x_{+}^{*}=x_{-,}\: k^{*}=k^{-1} \ .
\end{equation}
Indeed it is very simple, to show that for $|q|=1$ the Hopf star
$x_{+}^{*}=x_{+,}x_{-}^{*}=x_{-,}\: k^{*}=k$ can not be implemented
in this example. This is so since there is no representation of the algebra
$U_q(sl(2))$ by $2\times 2$ hermitian matrices. Recall that we have a
positive definite inner product (the one associated to the quantum version
of this model), in the vector space where the linear operators $x_+, x_-$
and $k$ act.

%%%%%%%%%%%%%%%%%%%%%%%%%%%%%%%%%%%%%%%%%%%%%%%%%%%%%%%%%%%%%%%%%%%%%%%%%%

\section{Concluding remarks}

The star structures of Hopf algebras appearing in physics depend on the
model, and are not necessarily Hopf stars. We have seen that for
factorizable scattering models and quantum mechanics of QSC there is a
compatibility with the Hopf structure. However, for statistical vertex
models and statistical physics of QSC this is not the case, and a twisted
star is obtained. As we already mentioned, this difference can be traced
back to the Wick rotation \cite{Itz} that connects quantum mechanics with
statistical physics\footnote{
In brief, consider the following operator for a quantum mechanical
system: $U(z)=\exp{-zH}$ where $H$ is the Hamiltonian of the system
(assumed to be time independent) and $z$ a complex number. If you restrict
$z$ to the imaginary axis, then $U(z)$ is the quantum mechanical evolution
operator. If you ``rotate'' (Wick rotation) the $z$ variable to the
positive real axis you get the central object of quantum statistical
mechanics, the quantum Boltzmann weight $Z = \exp{-\beta H}$.
}.
Twisted stars have different properties than Hopf stars \cite{cgt}; in
particular they do not form a tensorial category with respect to the usual
tensor product. However, the present analysis shows their physical relevance
in the field of integrable models.

%%%%%%%%%%%%%%%%%%%%%%%%%%%%%%%%%%%%%%%%%%%%%%%%%%%%%%%%%%%%%%%%%%%%%%%%%%

\bibliographystyle{amsalpha}

\end{document}